\documentclass[aps,pra,preprint,superscriptaddress,amssymb,amsfonts,showpacs]{revtex4}

\usepackage{color}
\usepackage[dvips]{graphics,graphicx}

\usepackage{calc,ifthen}

\newlength{\elimdepthdim}
\newlength{\elimheightdim}
\newlength{\elimwidthdim}

\newlength{\strutdepthdim}
\newlength{\strutheightdim}
\newlength{\strutwidthdim}

\newcommand{\ket}[1]{\qvbar{#1}\qrangle}
\newcommand{\bra}[1]{\qlangle{#1}\qvbar}

\def\id{{\mathchoice {\rm 1\mskip-4mu l} {\rm 1\mskip-4mu l} {\rm
1\mskip-4.5mu l} {\rm 1\mskip-5mu l}}}

\newcommand{\proof}{\paragraph*{Proof.}}

\newcounter{herefignum}

\newcommand{\shortqph}[1]{}

\providecommand{\ignore}[1]{}

\def\C{{\mathbb{C}}}

\def\proof{{\em Proof:  }}

\def\openone{\leavevmode\hbox{\small1\kern-3.8pt\normalsize1}}
\def\RR{{\rm I\kern-.2emR}}
\def\tr{{\rm tr}\; }
\def\trq{{\rm tr_Q} }
\def\0{{\bf 0}}

\def\proof{\noindent {\bf Proof:  }}

\def\openone{\leavevmode\hbox{\small1\kern-3.8pt\normalsize1}}
\def\RR{{\rm I\kern-.2emR}}
\def\tr{{\rm tr}\; }

\def\ca{{\cal A}}

\def\id{{\rm id}}

\newcommand{\QED}{\hspace*{\fill}\mbox{\rule[0pt]{1.5ex}{1.5ex}}}
\providecommand{\ignore}[1]{}

\newcommand{\ceil}[1]{\lceil #1 \rceil}
\newcommand{\floor}[1]{\lfloor #1 \rfloor}

\renewcommand{\ket}[1]{| #1 \rangle}
\renewcommand{\bra}[1]{\langle #1 |}
\newcommand{\proj}[1]{\ket{#1}\! \bra{#1}}

\newcommand{\inner}[2]{ \langle #1 | #2 \rangle}
\newcommand{\melement}[3]{ \langle #1 | #2 | #3 \rangle}
\newcommand{\dmelement}[2]{ \langle #1 | #2 | #1 \rangle}
\newcommand{\bitem}{\begin{itemize}}
\newcommand{\eitem}{\end{itemize}}
\newcommand{\benum}{\begin{enumerate}}
\newcommand{\eenum}{\end{enumerate}}
\newcommand{\beq}{\begin{equation}}
\newcommand{\eeq}{\end{equation}}
\newcommand{\beqa}{\begin{eqnarray}}
\newcommand{\eeqa}{\end{eqnarray}}

\newtheorem{definition}{Definition}

\newtheorem{theorem}{Theorem}
\newtheorem{proposition}{Proposition}

\newtheorem{lemma}{Lemma}

\newcommand{\bproof}{\begin{proof}}
\newcommand{\eproof}{\end{proof}}
\newcommand{\bprop}{\begin{proposition}}

\newcommand{\bdef}{\begin{definition}}

\begin{document}


\title{Semidefinite programming characterization and spectral adversary method
for quantum complexity with noncommuting unitary queries}



\author{Howard Barnum}
\shortqph{\email[]{barnum@lanl.gov}}

\date{September 13, 2006}

\begin{abstract}
Generalizing earlier work characterizing the quantum query
complexity of computing a function of an unknown classical ``black box''
function
drawn from some set of such black box functions,  
we investigate a more general quantum query model in which 
the goal is to compute
functions of $N \times N$ ``black box'' unitary matrices drawn from 
a set of such matrices, a problem with
applications to determining properties of quantum physical systems.  
We characterize the existence of an algorithm for such a query problem, 
with given query and error, as equivalent to the feasibility of a certain set of semidefinite
programming constraints, or equivalently the infeasibility of a dual of these 
constraints, which we construct.  Relaxing the primal constraints to correspond 
to mere pairwise near-orthogonality of the final states of a quantum computer, conditional
on the various black-box inputs, rather than bounded-error distinguishability,
we obtain a relaxed primal program the feasibility of
whose dual still implies the nonexistence of a quantum algorithm.  We use this to obtain
a generalization, to our not-necessarily-commutative setting,
of the ``spectral adversary method'' for quantum query lower bounds. 
\end{abstract}



\maketitle

\section{Introduction}
Quantum computers can solve certain problems faster than any known
classical algorithm: the best-known examples are probably Shor's
algorithm \cite{Shor94a, Shor97a}
for factoring integers in time polynomial in the number of
digits needed to represent them, and Grover's ``search'' algorithm
\cite{Grover96a},
which, for example, allows quadratic speedup (from time of order $N$
to time of order $\sqrt{N}$) of ``brute-force search'' for solutions
to certain problems.  The structure of these algorithms may be
understood as based on ``black-box'' or ``query'' algorithms, in which
we have as input a function implemented as a ``black-box'' subroutine,
and we would like to determine a property of the black-box function
with few calls ({\em ``queries''}) to the subroutine.  For factoring, the corresponding
query algorithm is one in which, given a strictly periodic function as a
black-box, we must find its period\footnote{``Strict'' periodicity means 
that not only does $f$ take the same value when its input is shifted by the period, but
it takes distinct
values on distinct inputs not obtainable from each other by shifting by a multiple
of the period.  The situation is slightly more
complicated for Shor's algorithm because the function is in fact only approximately
strictly periodic, but this makes no essential difference.}; for Grover's, given a 0/1
valued function taking, say, $n$-bit strings as inputs, we must
determine if the function is identically zero or not.  In abstract models
of such black-box computation, called ``query models'' by computer
scientists, an instance of a problem is specified by a set of possible
black-box functions, and a property of those functions (whose value, 
in some finite set, may depend on the function), which we want to compute with
bounded error (say, less than some constant $\varepsilon$).  The {\em
query complexity} of an instance is the minimal number of queries
needed to compute the property on that instance.  The cost of
computation done between queries is ignored in this abstract model.
Typically we are concerned with a problem having arbitrarily large
instances, and with how the query complexity of instances scales with
their size--for instance, polynomially in the case of the
``order-finding'' query problem \cite{Cleve2000a}
on which Shor's algorithm is based,
exponentially (but with half the classical exponent) in some versions
of Grover's algorithm.  In concrete algorithms such as Shor's
factoring algorithm, or applications of Grover's algorithm to speeding
up the search for solutions to instances of hard problems, the black
box is replaced by an explicit program or circuit, usually
a polynomial time program or polynomial size circuit, 
but the algorithm treats it as a black-box, i.e. does
not look at details of the program or circuit, but only provides
inputs to it and processes outputs from it.  Also, in such concrete algorithms
based on black-box ones explicit algorithms
must be provided for the computation that takes place betwen the
queries, and this, too, is typically polynomial-time in input size.
If the abstract black-box complexity of a problem is polynomial, {\em
and} concrete algorithms can be founded implementing each black box
polynomially, and each inter-box computation polynomially, then the
abstract black box algorithm can be converted into a concrete
polynomial-time algorithm, as in the case of factoring.  Lower bounds on 
black-box algorithms can imply lower bounds on the performance of 
concrete algorithms having such a substituted-black-box structure, but 
for these to be interesting, the possibility of known, easy ways of 
exploiting the structure of circuits in a concrete algorithm must be 
built in, for example by applying a lower bound technique to a set of
queries  including the inverse of the basic black-box transformations,
if the circuit model allows (as does the standard quantum one) the easy
construction of a polynomial-size circuit for the inverse of a given 
polynomial-size circuit.  Likewise, the ability to apply a black box 
or not conditional on the value of some qubit should also be included 
for similar reasons (given a quantum circuit, it is easy to concoct 
another circuit of essentially the same size that applies the first 
conditionally).  (This, and the point about the inverse, was suggested 
to me by Daniel Gottesman \cite{Gottesman2005a} 
at a talk I gave on an earlier version of this paper.)

In Grover's algorithm, and many other abstract query algorithms such
as the ``Abelian hidden subgroup'' problem that can be abstracted from
Shor's algorithm and its predecessors such as Simon's algorithm
\cite{Simon97a}, the black-boxes can be viewed as a set of commuting
unitaries implementing ``black-box functions'' quantum-coherently.
For example, they may compute a Boolean function $f: \{1,..,S\}
\mapsto \{0,1\}$ of an input in $\{1,...,S\}$ supplied in an
$S$-dimensional quantum register in its ``standard'' or
``computational'' orthonormal basis $\ket{i}, i \in \{1,...,S\}$, and
then write the resulting value onto an output qubit by adding it
modulo 2 to the value of the output qubit in its standard basis
$\{\ket{0}, \ket{1}\}$; thus $O_f \ket{i}\ket{b} = \ket{i} \ket{f(i)
\oplus b}$, for standard orthonormal bases of the two registers.  For
all the various possible such $f$, these ``black-box'' unitaries $O_f$
commute with each other, being diagonal in the basis that is the
product of the standard bases.  Obviously, one can do something
similar for a larger finite set of outputs.  Other models for quantum
queries to classical functions, such as ``phase queries,'' $O_f
\ket{i} \mapsto (-1)^{f(i)}\ket{i}$, equivalent up to a constant
factor in the number of queries to the above straightforward
quantum-coherent reversible computation of $f$ when conditioning and
the adjoint unitaries are included, are sometimes used, and there too
all unitaries commute.

In this paper, however, we analyze the case where queries involve a
not-necessarily commuting set of black-box unitaries.  This latter
setting is relevant, for example, to algorithms intended to extract
information about quantum physical systems, an area of intensive
research.  Although not all unitaries (e.g. on $n$ qubits, a
$2^n$-dimensional quantum system) can be represented with polynomially
many (in e.g. $n$) quantum gates, the ones that can are still of great
interest.  Many interesting questions about unitaries are still
superpolynomially hard (relative to P $\neq$ NP) when confined to such
polynomially representable unitaries.  Thus, just as in the case of
``quantum-coherent classical'' queries, there is the possibility that
abstract query algorithms for determining properties of noncommuting
quantum black-boxes may lead to efficient and important concrete
algorithms.
 
Note, for example, that unitary evolutions induced by ``local''
hamiltonians on a lattice can be well approximated by polynomially
many gates \cite{Lloyd96b}.  To extract certain information
(e.g. about the spectrum) directly from the unitaries themselves
involves manipulating $2^n \times 2^n$ matrices.  One could imagine
using the short classical description of the small quantum circuit
directly (i.e. in some way other than running the circuit, thereby
going beyond the black-box model) to do the computation more quickly,
even classically, but it is not clear that this will be possible and
for certain problems, it is not possible in polynomial time unless P =
NP.  However, there is the tantalizing possibility that least some
information may be gotten more efficiently than classically by
treating the unitary as a ``black box'' in a quantum computation
(legitimate in terms of actual computation time when it has a
poly-size quantum circuit).  Important candidate examples where the
quantum algorithm is better than known classical ones include
\cite{Knill1998a}, \cite{Poulin2003a}, \cite{Emerson2004a},
\cite{Poulin2004a}.

An important part of the study of quantum computation has been the
investigation of lower bounds on the quantum query complexity of
various problems.  Although lower bounds in query settings do not
logically imply lower bounds of the same functional form for concrete
versions of corresponding problems, because of the possibility of
``looking inside the black box'' in a concrete situation, many
computer scientists view them as a good guide in many situations: for
example, the lower bounds on Grover's problem \cite{Bennett97b}
matching the $\sqrt{N}$ performance of Grover's algorithm are widely
taken as fairly reasonable grounds to expect that quantum computers will not
perform NP-hard computations in polynomial time, although they are only
part of the story as a crucial part of the question is whether one believes
quantum circuits encoding classical computations may have some structure that 
quantum algorithms can take advantage of better than it is generally thought
classical computations can take advantage of classical circuit structure.

In this paper, we provide a new formulation of the quantum query
computation model with unitary black-box queries.  It closely parallels
the formulation for quantum-coherent classical queries in \cite{BSS2003};
all of the results in this paper have counterparts there and many of 
the ideas used in their proofs are related (indeed some parts are 
essentially identical) as well.
As for  quantum-coherent classical queries, our formulation takes the form
of a theorem showing that a query algorithm for a problem instance
exists if, and only if, a feasible solution to a certain set of
semidefinite programming (SDP) constraints exists.  This formulation
contains, we think, the mathematical ``essence'' of quantum query
complexity: much information concerning details of the unitaries
implementing the between-query evolution in the standard picture, but
irrelevant to the algorithm's query complexity, is not present in our
picture.  The formulation allows us to derive space bounds for unitary
query computations.  It allows us to exploit the ``revolution'' of the
last 15 years or so in conic, especially semidefinite, programming,
leading to polynomial-time methods for solving these optimization
problems, to obtain a polynomial algorithm for estimating the quantum
query complexity of a problem instance.  

\section{Mathematical and notational preliminaries}

In this next section, we will formalize two equivalent notions of quantum
query algorithm, and use them to formally define quantum query
complexity.  First, however, we record some mathematical conventions,
terminology, and facts we will use.  We often define a set $S$ as the set of
all things referred to by some expression $Expr(X)$ containing a
variable, as the variable ranges over another set, say $T$; we write
this as: $S := \{Expr(x)\}_{x \in T}$.  The pure states of quantum
systems, which are vectors in a complex inner product space of finite
dimension $d$ (we'll sometimes refer to it as a Hilbert space), will
often be identified, usually without comment, with the isomorphic
linear space of $d \times 1$ matrices (``column vectors'') over $\C$,
where the matrix is identified with the matrix elements of the state
in some special basis.  This special basis will the basis used to define operators
on that space.  Thus the space of operators on a quantum space will
also usually be implicitly identified with a space of matrices, and
states, both pure and mixed, on tensor products of quantum spaces will
also be identified with spaces of matrices, whose entries are interpreted as the
operators' matrix elements in the product of the
standard bases for the individual spaces.  Dirac notation will
sometimes, but not exclusively, be used for vectors, or for projection
operators, when especially when these represent, or directly correspond to, 
quantum states of a query computer.  We write $M(d)$ for the space of $d \times d$ complex
Hermitian matrices.  

The notion of ``purification'' of a mixed state $\rho^{H_1}$ (operator on 
a Hilbert space $H_1$) will also be used.  This is a ``pure''
state $\ket{\Psi^{H_1H_2}} \in H_1 \otimes H_2$ such that 
$\tr_{H_2} \proj{\Psi^{H_1 H_2}} = \rho^{H_1}$.  
We write $d_i$ for the dimension of $H_i$, $i \in \{1,2\}$.  
It is a well known fact that
{\em any} finite-dimensional positive semidefinite matrix $\rho$ on $H_1$
has a ``purification'' in $H_1 \otimes H_2$ as long as $d_2$ is 
at least $rank(\rho)$.  A sometimes useful way of thinking about states
on tensor products of spaces, and the partial 
trace, is in a block matrix picture:  identifying the space of operators on 
$H_1 \otimes H_2 $ with the
space of $d_1 \times d_1$ block matrices with blocks in $M(d_2)$
(viewed as arrays of matrix elements of the operator in the tensor 
product of standard bases for $H_1, H_2$).  Then the partial trace 
over $H_1$ of a matrix $M$ is the sum of its diagonal blocks, whereas 
its partial trace over $H_2$ is the matrix of traces of its blocks.
Parenthetical superscripts, like $G^{(X,Y)}$, indicate blocks of a 
block matrix.   

Superscripts are used (as we have just done) to denote which system an
operator acts on, or a vector belongs to (in the latter case they
occur within the ket or bra notation), subscripts to index vectors or
matrices belonging to an indexed set of such objects.  ``Functional''
notation like $\ket{\Psi(t)}$, $\rho(t)$ also indicates dependence on
an index, but its use will be confined to quantum states and variables
directly related to quantum states, such as the variables in the
``primal'' semidefinite programs we define below, that correspond
closely to quantum algorithms.  The reason for
this is that occasionally we want a quantum state to depend on which
black box has been supplied to a quantum algorithm as ``input'', and
we reserve subscripts, as in $\ket{\Psi_X}$, to indicate this
dependence on an an input $X$.  We often write, for example, the $X,Y$
matrix element of $G$ as $G[X,Y]$; when an object is a quantum state,
Dirac notation such as $\bra{X} G \ket{Y}$ may be used as well.
Because of the other uses to which we put subscripts, they are never
used to indicate matrix elements.

We use the notation $|S|$ to indicate the cardinality of a set $S$.  
When a Hilbert space is defined in terms of a distinguished orthonormal
basis indexed by a set $S$ (i.e., defined as the free complex inner product
space over the set $S$), we may also use $S$ to refer to the Hilbert 
space itself.  Quite generally, we also write $|S|$ for the dimension of
a Hilbert space $S$; there it does {\em not}, of  course, refer to its cardinality.

\section{Formulation of quantum query algorithms and complexity}

We will use both a ``black-box'' and an equivalent ``explicit input''
model of quantum query complexity.  In the black box model, a problem
is given by specifying a set $S$ of ``black-box'' unitary operators, a
finite set $T$, and a function $g: S \rightarrow T$.  The problem is
to design an algorithm that, for all $X \in S$, computes $g(X)$,
exactly or with zero or bounded error.  We will mostly be interested
in the bounded error case.  The computer state will be written as a
superposition of basis vectors $\ket{w}\ket{i} \in Q \otimes W$, where
the first, $n$-dimensional, register $Q$ is the ``query register'', on
which the unitary $U \in S$ acts, and the second register, $W$, is
workspace.

For what follows, we will let $S$ be a finite set of unitaries in
order to avoid having ``matrices'' indexed by infinite sets, or
operators on infinite-dimensional spaces, though we expect
generalizations to infinite sets of unitaries to be straightforward.

\begin{definition} 
A {\em finitary query problem instance} in the unitary-queries model 
({\em problem} for short) is 
an integer $n$, a finite set $S$ of $n \times n$ unitaries, a finite
set $T$, and a function $g: S \rightarrow T$.
\end{definition}

\begin{definition}
A $q$-query
quantum algorithm (QQA) for a problem $P = (n, S, T, g)$
is an integer $|W|$ (the ``workspace dimension''), a sequence $U_0, U_1, ...,
U_{q}$ of $n|W| \times n|W|$ unitary matrices (the ``inter-query unitaries''), 
and an indexed set of $|T|$ projectors $\{P_z\}_{z \in T}$, that are
$n|W| \times n|W|$ matrices.  
\end{definition}

On a black-box
unitary input $X$, such an algorithm runs as follows.  We consider a computer
whose Hilbert space is $Q \otimes W$, the tensor product of an $n$-dimensional
``query register'' $Q$ which has a distinguished orthonormal basis indexed by 
$0, ..., n-1$, and a $|W|$-dimensional ``workspace'' $W$ with a distinguished
basis $0, ..., |W|-1$.  We define an action of 
the unitary matrices $U_i$ on this space
by interpreting them as the matrices of unitary operators $Q \otimes W$
in an ordered basis $\ket{i}\ket{j}$ (with the fast running index 
corresponding to $Q$).
We start with the computer state
$\ket{0}\ket{0}$, and alternate
the unitaries $U_i$ with the fixed query unitary $X$ (which acts only on the 
register $Q$, i.e. we apply $X \otimes I$ to the computer).  Thus at 
time $t$ (immediately after the $t$-th query) 
the state of the computer when $X$ is input is:
$\ket{\phi_X(q)} = 
U_t (X \otimes I) U_{t-1} (X \otimes I) \cdots U_1 (X \otimes I) U_0\ket{0}\ket{0}$.  
After $q$ queries,
the projectors $\{P_z\}_z$ are measured, obtaining an outcome $z \in
T$, interpreted as the value of $g(X)$, with probability \beq
p(z) = \dmelement{\Phi_X(q)}{P_z}\;.  \eeq

The special case of computing a Boolean function $f: \{0,1\}^m 
\mapsto \{0,1\}$ using phase queries corresponds to the 
commuting set of unitaries $S = \{U_x\}_{x \in \{0,1\}^m}$, defined by 
$\melement{i}{U_x}{j} = \delta_{ij} (-1)^{x_i}$.

\begin{definition} \label{def: QQA computes g}
We say an algorithm $A = (|W|, \{U_0, ... U_{q-1}\}, \{P_z\}_{z \in T})$
solves the problem $P := (n, S, T, g)$
with bounded error $\varepsilon$ (or for short, $\varepsilon$-computes $g$), 
iff with $\ket{\phi_X(q)}$ defined as above, for all $z \in T$ and 
$X \in S$ such that $g(X)=z$,
\beq \label{QQA output condition}
\dmelement{\Phi_X(q)}{P_z} \ge \varepsilon\;.
\eeq
\end{definition}
We sometimes call such a QQA a ``$(q,\varepsilon)$-QQA for $g$''.

\begin{definition} \label{def: quantum query complexity}
The {\em quantum query complexity} $QQC_\epsilon(g)$ of a function 
$g$ is the least integer $q$ such that there exists a $q, \varepsilon$-QQA
for $g$.  
\end{definition}

At times it will be useful to consider an ``extended computer'' whose 
Hilbert space is $I \otimes Q \otimes W$, with $Q$, $W$ as before and 
$I$ an $|S|$-dimensional ``input'' register with a distinguished orthonormal basis
$\{\ket{X}\}_{X \in S}$.  With such a construction, we can give an extended
``explicit input'' version of quantum query algorithms.  The matrices 
$X$ acting on $Q$ are replaced by a single unitary matrix $\Omega$ acting on 
$I \otimes Q$ by ``reading the input out of I in the standard basis''
and, conditional on reading input $X$, doing the unitary $X$ on 
the register $Q$.  That is, in the tensor product basis $\ket{X}\ket{i}$,
$\Omega$ acts via: 
\beq
\Omega \ket{X}\ket{i} = \ket{X} X\ket{i}\;.
\eeq
Thus the matrix $\Omega$ written in this basis, with $Q$'s basis 
the fast-running
index, is block-diagonal, with the unitaries $X$ as the diagonal blocks.

We can view the first $t$ steps of an algorithm as acting on such a
computer (starting in an initial state $\ket{\Psi^{IQW}}$) to produce a
state $\ket{\Psi^{IQW}(t)}$ defined as follows:
\begin{definition} \label{def: state of extended computer at t}
\beq \label{eq: state of extended computer at t}
\ket{\Psi^{IQW}(t)} := (I^I \otimes U_t^{QW})(\Omega^{IQ} \otimes I^W) 
(I^I \otimes U^{QW}_{t-1}) (\Omega^{IQ} \otimes I^W) 
\cdots (I^I \otimes U^{QW}_1) (\Omega^{IQ} \otimes I^W) (I^I \otimes U^{QW}_0)
\ket{\Psi^{IQW}}\;.  
\eeq
\end{definition}
Here we have introduced superscripts on unitaries to indicate which systems they act on,
and superscripts inside kets to indicate the systems they belong to.
These are not always used, however; sometimes we let the context make it clear what
an operator acts on.
Notice that the queries $\Omega$
do not touch the workspace (and only touch the input
register to read it in the standard basis), while the inter-query unitaries
may arbitrarily entangle $Q$ and $W$, but do not touch the ``notional'' input
register $I$.

As we will see in the proof of the main theorem,
some of the variables in the semidefinite program 
we will now define, and which appears in our first main theorem characterizing
query complexity, can be interpreted as the density matrices of the subsystems 
$I \otimes Q$ or $I$ of such an extended query computer whose query and work registers
are started in $\ket{0}\ket{0}$, and whose input register is started in an 
unnormalized equal superposition of inputs (so that
$\ket{\Psi^{IQW}} = \sum_{X \in S} \ket{X}\ket{0}\ket{0}$).

\section{SDP characterization of quantum query complexity: primal formulation}

\begin{definition}[Semidefinite program $P(g,q,\varepsilon)$]
\label{def: primal}
By $P(g,q, \varepsilon)$ we mean the following semidefinite program feasibility
problem:
Find $|S|n \times |S|n$ positive semidefinite Hermitian matrices
$\rho^{IQ}(t), ~t \in \{0,...,q-1\}$, an $|S| \times |S|$ PSD Hermitian 
matrix $\rho^I(q)$ and $|S| \times |S|$ PSD matrices
$\Gamma_z, {\rm ~for~all~} z \in T$, satisfying the constraints:
\beqa \label{initial  unitary constraint}
\trq \rho^{IQ}(0) = E \\
\tr_Q \rho^{IQ}(t) = \tr_Q \Omega \rho^{IQ}(t-1) \Omega^\dagger  
\label{first computation constraints}
\eeqa
for $t \in \{0,..., q-1\}$, 
where
$E$ is the constant all-ones matrix),
\beqa \label{final computation constraint}
\rho^I(q) = \tr_Q \Omega \rho^{IQ}(q-1) \Omega^\dagger\;, \\
\label{output1}\sum_{z \in T} \Gamma_z &=& \rho^{I}(q)
\\
\label{output2}\Delta_z*\Gamma_z & \succeq & (1-\varepsilon)\Delta_z,
\eeqa
where the constant diagonal matrix $\Delta_z$ is defined by $\Delta_z(X,X)=1$
if $g(X)=z$, else $0$.  $*$ denotes the elementwise (aka Schur or Hadamard)
product of matrices.
\end{definition}

Using this, we state the following theorem, which is the first main result
of the paper:

\begin{theorem} \label{theorem: main characterization theorem}
A $q$-query, $\varepsilon$-error quantum algorithm to compute
$g: S \rightarrow T$ exists if and only if a feasible solution 
to $P(g, q, \varepsilon)$ does.  Furthermore, for each particular
feasible solution 
$\left\langle  \{\rho^{IQ}(t)\}_t, \rho^I(q), \{\Gamma_z\}_{z \in T}\right\rangle$ 
there is a $(q, \varepsilon)$-$QQA$ that computes 
$g$, for which the
dimension $r$ of the working memory is no larger than the greater
of of $|S|N$ and $\ceil{\sum_{z \in T} rank(\Gamma_Z)/N}$.  Since the 
latter is no greater than $\ceil{|S||T|/N}$, it follows that any
$(q,\varepsilon)$-$QQA$ computing $g$ 
may be implemented with workspace dimension no greater than
$\max{|S|N, \ceil{|S||T|/N}}$ in addition to the $N$-dimensional query 
register.  
\end{theorem}

In terms of qubits,  then, the algorithm needs
no more than $\max{ \left\{ \ceil{\log{|S|}} + \ceil{\log{N}}, 
\ceil{\log{S}} + \ceil{\log{|T|}\right\} }
- \floor{\log{N}} }$ 
qubits of workspace in addition to the $\ceil{\log{N}}$-qubit
query register.  

\noindent
{\bf Proof:} We prove first the implication from the existence of a
$(q,\varepsilon)$-QQA solving the problem to the existence of a feasible
solution to $P(g,q, \varepsilon)$, establishing it by constructing the
latter from the former.  We do this by defining matrices $\rho^{IQ}(t)$,
$\Gamma_z$ in terms of the objects of the QQA, and showing that they
satisfy the constraints (\ref{initial unitary constraint}--\ref{output2}) 
on the variables of the same names in the definition 
of $P(g,q, \varepsilon)$.

We begin by showing that in order to tell whether an algorithm 
will succeed in $\varepsilon$-computing
the function $g$ no matter what the input, {\em all} we 
need to know is whether the geometry (the inner products) of the
final 
computer states $\ket{\Phi^{QW}_X(q)}$ 
allows these states to, roughly (i.e. up to $\varepsilon$), 
lie in  a set of 
orthogonal subspaces such that the vectors $\ket{\Phi^{QW}_X}$
in each subspace share
the same value of $g(X)$.  (They may have to be isometrically 
embedded in a larger space to do this.)  Formally, this gives an SDP which we 
now construct.  
We array the inner products in
a matrix $G(q)$ defined 
\begin{definition} \label{def: gram matrix M}
$\tilde{M}(q)[X,Y] := 
\inner{\Phi^{QW}_X(q)}{\Phi^{QW}_Y(q)}$.
\end{definition}   
For later use, 
similar matrices $G(t)$ may be defined for
all $t$ between $0$ and $q$ inclusive, using the conditional computer
states $\ket{\Psi^{IQW}(t)}$ after the $t$-th query and post-query 
unitary.  The $t=0$ case, before any query, is 
of course the all-ones matrix.  Because these are matrices of inner
products (sometimes called ``Gram matrices''), they are necessarily
positive semidefinite.  The condition that the geometry of the
final inner products is correct may be stated as a semidefinite programming
feasibility problem with a constraint involving $G(q)$:

\begin{definition}[SDP $O(g, \varepsilon, M)$] \label{def: output SDP}
For a problem $g$, real number $\varepsilon$ between zero and one,
and $|S| \times |S|$ positive semidefinite matrix $M$, 
the program $O(g, \varepsilon, M)$ is the following:
Find $|S| \times |S|$ PSD matrices $\{\Gamma_z\}_{z \in T}$ such that 
\beqa 
\sum_{z \in T} \Gamma_z & = & G \label{O1}\\
\Delta_z * \Gamma_z & \succeq & (1-\varepsilon)\Delta_z.\label{O2}
\eeqa
\end{definition}

The proof of the following lemma essentially repeats part of the
proof of the main theorem in \cite{BSS2003}.

\begin{lemma} \label{lemma: QQA output implies feasibility of output SDP}
The SDP $O(g, \varepsilon, G(q))$, where $G(q)$ is defined as in 
Definition \ref{def: gram matrix M} above to 
be the final-state inner-products matrix of a QQA for $g$, is feasible
if the QQA $\varepsilon$-computes $g$.
\end{lemma}

\noindent
{\bf Proof of lemma:} The feasible solution is obtained by defining 
$\Gamma_z$ as the matrices with components:
\beq \label{def: feasible gammaz}
\Gamma_z[X,Y] :=  
\dmelement{\Phi^{QW}_X(q)}{P_z}\;.
\eeq
Satisfaction of the constraint (\ref{O1})
follows because $\sum_z P_z = I$, while (\ref{O2}) 
is guaranteed by 
Eq. (\ref{QQA output condition}) in Definition \ref{def: QQA computes g}.
\QED

The definition of $\Gamma_z$ just given is also the one will use to 
show feasibility of $P(q, \varepsilon)$.

Lemma \ref{lemma: QQA output implies feasibility of output SDP}
 has a suitable converse (see below).  Thus to decide, from
the final inner-products matrix $G(q)$, whether the value of
$g$ has been $\varepsilon$-computed or not, is a question of semidefinite
program feasibility.  However, essentially because the action of the
queries is not linear on the matrices $G(t)$ that we defined
based on the QQA (the inner-product matrices of the input-conditioned
states after query $t$), we cannot formulate linear constraints on
variables corresponding to $G(t)$ that enforce the condition
that the final inner-products matrix must arise from the initial one
via queries and pre- and post-query unitaries.  We need different, though
related, quantities to fomulate that condition as a linear constraint.

These quantities are most easily and intuitively described by going to
the ``explicit inputs'' formulation described above, with overall
state space $IQW$ including the ``virtual input register'' $I$ started
in an unnormalized uniform superposition of inputs.  It is easily
seen, using the definitions of $\Omega$, $\ket{\Psi^{QW}_X(t)}$, and
$\ket{\Psi^{IQW}(t)}$, that
\beq
\ket{\Psi^{IQW}(t)} = \sum_{X \in S}\ket{X^I} \ket{\Phi^{QW}_X(t)}\;.
\eeq
We define $\rho^{IQW}(t) := \proj{\Psi^{IQW}(t)}$, and density matrices
such as $\rho^{IQ}(t):= \tr_W \rho^{IQW}(t)$, etc....  It is then easily
seen by direct calculation that
\beqa \label{eq: final-states gram matrix is input register density matrix}
\bra{X}\rho^{I}(t)\ket{Y} = \inner{\Phi_X(t)}{\Phi_Y(t)}\;,
\eeqa
and consequently that the matrix of $\rho^I(t)$ in the standard basis
that labels inputs, is just the Gram matrix $G(t)$ of Definition 
\ref{def: gram matrix M}.  We will generally
identify operators with their matrices in the standard tensor product
basis for $IQW$, and hence if the QQA $\varepsilon$-computes $g$, the
program $O(g, \varepsilon, \rho^I(q))$ with $\rho^I(q) := \tr_Q[\rho^{IQ}(q)]$
in place of $M$, is feasible.  Moreover, the quantities $\rho^{IQ}(t)$
are exactly those necessary to formulate the computational constraints
linearly, as we now show.

Since in our analysis we will at times
consider separately the effects of the query and of the post-query 
unitary, we also define $\ket{\Phi^{QW}_X(t+)} := 
(X^Q \otimes I^W) \ket{\Phi^{QW}_X(t)}$, 
$\ket{\Phi^{IQW}(t+)} := (\Omega^{IQ} \otimes I^W )\ket{\Phi^{IQW}(t)}$, 
and
$\rho^{IQ}(t+)$ as the ``density'' matrix 
$\tr_W \left[  \proj{\Psi^{IQW}(t+)} \right]$; these
are the vectors and density matrix after the $t$-th query but before the $t$-th post-query
unitary.  Since the post-query unitary $U^{QW}(t)$ does not touch $I$, 
$\rho^I(t) = \rho^I((t-1)+)$, or in other words:
\beq
\tr_Q \rho^{IQ}(t) = \tr_Q \rho^{IQ}((t-1)+)\;.
\eeq
Since the query is just the implementation of
the unitary $\Omega$ on $IQ$, we have:
\beq
\rho^{IQ}((t-1)+) = \Omega \rho^{IQ}(t-1) \Omega^\dagger \;.
\eeq
Eliminating the unnecessary quantities $\rho^{IQ}((t-1)+)$, we can
combine the two preceding sets of equations into a single set (indexed
by $t$) of linear equations:
\beq
\tr_Q  \rho^{IQ}(t) = \tr_Q \Omega \rho^{IQ}(t-1) \Omega^\dagger.
\eeq
In other words, the quantities $\rho^{IQ}(t)$ satisfy the constraints
(\ref{first computation constraints}).  It is also clear that 
$\rho^{IQ}(0)$ as defined from the algorithm satisfies 
(\ref{initial unitary constraint}), because $U^{QW}(0)$ does not touch $I$, and
$\ket{\Psi^{IQW}}$ has the all-ones matrix as its reduced density matrix.
Furthermore, since 
as stated in Eq. (\ref{eq: final-states gram matrix is input register density matrix}), 
$G(t) \equiv \rho^{I}(t)$ and the latter is just $\tr_Q \rho^{IQ}(t)$, we have from
Lemma \ref{lemma: QQA output implies feasibility of output SDP} and its
proof that $\Gamma_z$ as defined in that proof
satisfy the constraints
(\ref{output1}) and (\ref{output2}).  Thus we have shown the first direction of 
the theorem (existence of a QQA implies feasible solution to the SDP).

It remains to show the other direction, that the existence of a
feasible solution for $P(g,q, \varepsilon)$ implies that of an
$\varepsilon$-QQA solving the problem with the stated amount of
workspace.  Again it is a straightforward construction, though we must
keep track of the amount of workspace used in the algorithm we
construct.  In this part of the proof 
$\rho^{IQ}(t)$, $\Gamma_z$ will be taken to be the feasible
values of the variables of the same names in Definition \ref{def: primal};
it will turn out, of course, that when we have constructed the desired QQA, 
they will coincide with the quantities of the same names, 
$\rho^{IQ}(t)$, $\Gamma_z$, obtainable from that QQA via the definitions
in the first part of our proof.

The construction begins with a converse of Lemma 
\ref{lemma: QQA output implies feasibility of output SDP}.
\begin{lemma} \label{lemma: output SDP feasibility implies existence of final-state vectors}
If the SDP $O(g, \varepsilon, M)$, has feasible solution 
$\{\Gamma_z\}_{z \in T}$ there 
exists a set of vectors $\{\ket{\Psi_X}\}_{X \in T}$ in a Hilbert
space of dimension no greater than  $\sum_{z \in T} rank(\Gamma_z)
\le |S||T|$ and projectors $P_z$ acting on that space such that $M$ is the Gram matrix
of $\{\ket{\Psi_X}\}_{X \in T}$ and $P_z$ satisfy Eqs. (\ref{def: feasible gammaz})
and (\ref{QQA output condition}).
\end{lemma}

\noindent
{\bf Sketch of proof of Lemma 
\ref{lemma: output SDP feasibility implies existence of final-state vectors}:} 
The proof (with notational differences) may be
found in \cite{BSS2003}; it proceeds by constructing vectors
$\ket{\Theta_X}$ of length $|S|$ and a ``POVM'' consisting 
of $|S| \times |S|$ PSD matrices $\{R_z\}_{z \in T}$ such that $\sum_{z \in T}
R_z = I$ and $\dmelement{\Theta_X }{R_z} \ge \varepsilon$, and
then Naimark-extending the POVM to a set of projectors $P_z$ 
in a larger
space and identifying $\ket{\Psi_X}$ as the corresponding embeddings
of the vectors $\ket{\Theta_X}$ in the larger space.  This 
ensures that $\ket{\Psi_X}$ satisfy  (\ref{QQA output condition}).  The minimal
dimension required for the Naimark extension is $\sum_{z \in T}
rank(\Gamma_z)$. \QED

Since Eqs. (\ref{output1}) and (\ref{output2}) just state that  
$O(q, \varepsilon, M)$ with $\tr_Q \rho^{IQ}(q)$ substituted for $M$ is satisfied, 
Lemma \ref{lemma: output SDP feasibility implies existence of final-state vectors}
gives us vectors 
$\{\ket{\Psi_X(q)}\}_{X \in T}$ in a Hilbert space $H$ of dimension $|H| := 
\sum_{z \in T} rank(\Gamma_z)$ whose Gram matrix is $\tr_Q(\rho^{IQ}(q))$
and projectors $P_z$ on that space, which together satisfy Eqs. 
(\ref{def: feasible gammaz}) and (\ref{QQA output condition}).
We may give $H$ the structure $Q \otimes W$ with $Q$ 
$|S|$-dimensional and the dimension of $W$ large enough to guarantee that 
$dim(Q \otimes W) \ge \sum_{z \in T} rank(\Gamma_z)$; $|W| = 
\ceil{|H|/|N|} \le \ceil{|S||T|/|N|}$ suffices.
Given the vectors $\ket{\Psi_X(q)} \in Q \otimes W$, we can construct the
state $\ket{\Psi^{IQW}} := \sum_X \ket{X^I}\ket{\Psi^{QW}_X(q)}$.  By construction,
this state's reduced
density matrix for system $I$ will equal the feasible $\rho^I(q)$.

Now suppose we have $\ket{\Psi^{IQW}(t)}$ such that its reduced density matrix
coincides
with the feasible value $\rho^{IQ}(t)$ (or, for the case $t=q$, some arbitrary
$\rho^{IQ}(t)$ whose $I$ density matrix coincides with the feasible $\rho^I(q)$).
We construct 
$U^{QW}$ such that $\ket{\Psi^{IQW}(t-1)} := 
(\Omega^{IQ\dagger} \otimes I^W)(I^I \otimes U^{QW\dagger}) \ket{\Psi^{IQW}(t)}$, 
has $IQ$ reduced density matrix equal to  $\rho^{IQ}(t-1)$ (or, for the case
$t=1$, to the all-ones matrix).  To do this, first note
that any purification of 
$\Omega \rho^{IQ}(t-1) \Omega^{\dagger}$ into $W$ (and there exist many so long as 
$|W| \ge |I||Q|$) is also a purification
of $\rho^I(t) := \trq \rho^{IQ}(t)$ into $QW$, 
by the constraint (\ref{first computation constraints}).
Moreover, by acting via a unitary $U^{QW\dagger}$ 
on $\ket{\Psi^{IQW}(t)}$, we can reach such a purification 
of $\rho^{I}(t)$ that is also a purification of
$\Omega \rho^{IQ}(t-1) \Omega^{\dagger}$, as long as $W$
has dimension at least $|S|N$.  We let a $U^{QW}$ that achieves this 
be the $t$-th unitary, $U^{QW}(t)$ of our algorithm, and define 
$\ket{\Psi^{IQW}(t-1)} := 
(\Omega^{IQ\dagger} \otimes I^W)(I^I \otimes U^{QW\dagger}) \ket{\Psi^{IQW}(t)}$.
Thus, $\trq \proj{\Psi^{IQW}(t-1)} = \rho^{IQ}(t-1)$, as claimed.

We apply this step beginning with the states $\ket{\Psi^{IQW}(q)}$ already constructed,
until we get state $\ket{\Psi^{IQW}(0)}$ which by construction will have
the all-ones matrix as its reduced density matrix, and thus
$\ket{\Psi^{IQW}(0)} = \sum_X \ket{X} \ket{\chi^{QW}}$, where WLOG we can choose
$U^{QW}(0)$ so that $U^{QW^\dagger}(0)\ket{\chi^{QW}} = \ket{0}$.  Thus the sequence 
$U^{QW}(0),...,U^{QW}(q)$, and the indexed set $\{P_z\}_{z \in T}$ we have constructed
are a quantum algorithm that $\varepsilon$-computes $g$, and the dimension of $W$
satisfies the claimed bound, which derives from the bounds on $|Q \otimes W|$
of $\sum_z rank(\Gamma_z)$ 
(from the Naimark extension at the output)
and
$|S|N$ (from the workspace needed to reach an 
arbitrary purification of a fixed $\rho^{IQ}$ in the post-query unitary  step).  
\QED

\noindent
{\bf Remark:}  For those who like the matrix picture, 
thinking of the matrix $G(t)$ of $\rho^{IQ}(t)$ 
in the standard basis blocked according to $X$ and 
$Y$, we see that during the query each block is updated according to a fixed
block-dependent linear map:
\beq
G^{(X,Y)} \mapsto Y G X^\dagger\;.
\eeq
This is just conjugation by the block-diagonal
unitary matrix whose $(X,X)$ block is $X$ (i.e., the matrix of $\Omega$).

Using this we can express the constraints 
(\ref{first computation constraints}) in terms of the 
matrix $M$ viewed as blocked according to $X,Y$.  Each of the $q$ constraints
on the matrices $\rho^{IQ}(t)$
(which states that an 
$|S| \times |S|$ matrix calculated from $\rho^{IQ}$, 
namely its partial trace $\rho^{I}$, is equal to another such matrix),
becomes
$\frac{|S|(|S|+1)}{2}$ 
constraints each stating that the trace of an $(X,Y)$ block
of some matrix is equal to that of another:
\beq
\tr [ G^{(X,Y)}(t) ] = \tr [Y G^{(X,Y)}(t-1+) X^\dagger]\;,
\eeq
or, in the case $t=q$, a similar set of constraints with no trace on 
the LHS.
This is because the matrix of the partial trace in question is
the matrix of traces of the blocks; since the
block matrix is Hermitian, only $|S|(|S|+1)/2$ blocks, say 
those on and above the main diagonal of blocks, are independent.
Equivalently,
\beq
\tr [ G^{(X,Y)}(t)] = \tr [X^\dagger Y G^{(X,Y)}(t-1+)]\;.
\eeq
\QED

\section{The dual SDP}

In order to find the SDP feasibility problem dual to the one just
given, we begin by stating a very general theorem concerning feasibility
of conic program constraint sets.  

\begin{theorem} \label{theorem: general duality theorem}
Let $K$ be a closed, pointed, generating convex cone in an
$m$-dimensional real vector space $V$, with a distinguished inner
product $(\cdot ,\cdot )$.  Let $W$ be a $p$-dimensional real vector
space, also equipped with a distinguished inner product (written
similarly).  Let $K^* \subset V$ be the cone dual to $K$ according
$V$'s inner product.  Let $A$ be a fixed linear transformation from
$V$ to $W$ whose kernel is $\{0\}$, and let $b \in W$ be a constant
nonzero vector.  Let $A^*: W \rightarrow V$ be the linear map ``dual''
or ``adjoint'' to $A$, defined by $(w, Av) = (A^*w, v)$.  (For
example, if $V$ and $W$ are viewed as spaces of column vectors of
lengths $m$ and $p$ respectively equipped with the inner products
$(u,v) = u^t v$, and $A$ is represented by its $p \times m$ matrix
$\hat{A}$, then $A^*$'s matrix is $\hat{A}^t$.)

Consider the conic programming feasible set defined by:
\beq
P:=   \{ x \in V : Ax = b, x \ge_K 0 \} \;,
\eeq 

This set is empty (the constraints are 
``infeasible'') if and only if the dual feasible set
\beq
D :=  \{ y \in  W :  A^* y \ge_{K^*} 0, (b, y) <  0 \}
\eeq
is nonempty (the dual constraints are ``feasible'').
\end{theorem}

\proof First let $x$ belong to $P$.  Suppose that $A^* y  \in K^*$, so 
$y$ satisfies the first condition defining $D$.  We show that $(b,y) \ge 0$, 
so that $y \notin D$.  $A^*y \in K^*$ implies (since $x \in K$) that 
$(x, A^*y) \ge 0$.  Thus $(Ax, y) \ge 0$; since $x \in P$, $Ax=b$, so 
$(b,y) \ge 0$.  

Next, suppposing $P$ infeasible we construct a point in $D$.  
Consider the $A$-image of $K$, denoted $AK$.
By the assumption that $A$'s kernel is $\{0\}$, and for example Theorem 9.1
of \cite{Rockafellar70a}, $AK$ is a closed convex cone.  Now, $b \notin AK$, for if it
were, its preimage would belong to $P$, contradicting the supposition.  
Therefore, by (for example) Theorems 11.1, 11.3, and 11.7 of \cite{Rockafellar70a}, there 
exists a hyperplane through the origin properly separating $b$ and $AK$; this hyperplane
is the zero-set of a linear functional $L(x) := (y,x)$ determined by a vector
$y \in W$.  Thus (cf. the proof of Thm. 11.1 in \cite{Rockafellar70a}) $(b,y) < 0$,
and for all $z \in AK$, $(z, y) \ge 0$.  The latter is equivalent to: for 
all $x \in K$, $(Ax, y) \equiv (x, A^*y) \ge 0$.  Thus $y \in D$.  \QED

Lemma 2 of \cite{BSS2003} was a special case of this, for a particular
cone $K$ and a particular form of the linear map $A$.  
In \cite{BSS2003} we then further specialized the Lemma to 
the case in which the primal feasible set $P$ was the SDP characterizing
the existence of a quantum query algorithm for classical Boolean queries.
We now proceed by giving a generalization of Lemma 2 of \cite{BSS2003} 
which is still a special case of the above theorem, but which is sufficiently
general to encompass the SDP characterizing quantum query complexity with
arbitrary queries.  

\begin{lemma} \label{lemma: specialized duality lemma}
Let $K \subset W$ be the product of $k$ cones 
of PSD Hermitian matrices (with the $r$-th cone a cone of 
$d_r \times d_r$ matrices), in the obvious $W$ (direct sum of
the spaces of $d_r \times d_r$ Hermitian matrices).  Let $V$ be the direct sum of $k$ copies
of $H(s)$ for some fixed $s$.  Let ${\bf A}$ be a 
fixed $k \times k$ matrix whose entries are linear
maps $\ca_{\alpha, \beta}: H(d_\beta) \mapsto H(s)$.  
Let ${\bf B}$ be
a nonzero element of $V$, i.e. a $k$-tuple of matrices $B_\beta$, with 
$B_\beta \in H(s)$.
Equip $V$ and $W$ with the trace inner products, $(A, B) := \tr AB$.
(Matrices in  $V$ and $W$ are block-diagonal, $k$ blocks by $k$ blocks.)
Consider the ``primal'' feasible set:
\beq
P := \{ X \in W : \sum_\beta \ca_{\alpha \beta}(X_\beta) = B_{\alpha}, 
X \in K\}\;,
\eeq
and the ``dual'' feasible set
\beq
D := \{Y \in V: \sum_{\beta} \ca^*_{\beta \alpha}(Y_{\beta}) 
\ge_{K} 0\;, \sum_{\beta} \tr Y_\beta B_\beta < 0 \}\;.
\eeq 
Suppose further that the only feasible solution to $P_0$ (the primal problem with 
$B_\alpha$ set equal to zero) is $0 \in W$.  Then
if $D$ is feasible, $P$ is infeasible, and vice versa.
\end{lemma}

We caution the reader not to confuse the variable 
matrices $X_\alpha$, $Y_\beta$ 
appearing in the SDP above with the variables $X$ and $Y$ that we 
commonly let range over input unitaries in $S$. 
We will rarely use these notations 
together, and only when it is clear from the context what is meant, and in any case
we never use subscripts on the input unitaries, nor do we 
ever omit subscripts from 
the primal and dual variables of the above type of program.  

\noindent
{\bf Remark:}  Note that $\ca^*$ is the linear map often called by quantum
information theorists $\ca^\dagger$, defined 
by $\tr F \ca^\dagger(G) = \tr \ca(F) G$ (for all $F$ in  
the input space and $G$ in the output space, though 
it suffices to require it for bases of these spaces given linearity).  In the case where
$\ca$ is completely positive, i.e. $\ca: G \mapsto \sum_i A_i G A_i^\dagger$,
$\ca^\dagger$ may be defined via $\ca^\dagger: G \mapsto \sum_i A_i^\dagger
G A_i$. \QED

The program $P(g, \varepsilon, q)$ is a case of $P$, for which 
$W$ is the direct sum of $q$ copies of $H(|S|n)$ and $2|T|+1$ copies
of $H(|S|)$, and $V$ is the direct sum of $q + 2 + |T|$ copies 
of $H(|S|)$.  In terms of the associated query algorithm, 
the $q$ copies of $H(|S|n)$ in $W$ are where the density matrices
$\rho^{IQ}(t)$ will live, one of the copies of $H(|S|)$ is
for $\rho^I(q)$ and the other $2|T|$ copies of $H(|S|)$ are
for the the output conditions:  $|T|$ of them, 
indexed by $z \in T$,  for an 
additive decomposition of the final $\rho^{I}$ into positive 
matrices $\Gamma_z$
representing the portion of the output matrix for which the
final measurement has result $z$, and $|T|$ more for slack variable 
matrices $\Pi_z$, used to transform the inequality conditions on the
$\Gamma_z$, for succesful
computation, into equality conditions. 
These inequality conditions are are $\Delta_z * \Gamma_z \succeq
(1 - \varepsilon) \Gamma_z$;  requiring the slack
variables $\Pi_z$ to be positive while enforcing 
the equality constraint $\Delta_z * \Gamma_z - \Pi_z = (1 - \varepsilon)
\Delta_z$ is equivalent to imposing the inequality constraint
on the $\Gamma_z$.  Thus the vector of primal variables $X_\beta$
is indexed as follows:  for $0 \le \beta \le q-1$, $X_\beta 
= \rho^{IQ}(\beta)$; for $\beta = q$, $X_\beta = \rho^I(q)$; for 
$\beta = q+z (z \in T \equiv \{1,...,|T|\}),
X_\beta = \Gamma_z$; for
$\beta = q+1+ |T|+ z (z \in T \equiv \{1,...,|T|\}),
X_\beta = \Pi_z$.

We now specify the
maps $\ca_{\alpha,\beta}$ and constant vector ${\bf B} = [B_\alpha]$.
We will
give rows of the matrix $\ca_{\alpha, \beta}$, followed by the corresponding
RHS constant $B_\alpha$, since each row and $B_\alpha$ corresponds to 
a constraint;  the constraints will be naturally grouped by type.

For 
$0 \le \alpha \le q-1$, $\ca_{\alpha, \alpha}$ is the partial
trace map $G^{IQ} \mapsto \tr_Q(G^{IQ})$, and (for 
$1 \le \alpha \le q$) $\ca_{\alpha -1, \alpha}:
G^{IQ} \mapsto - \tr_Q(G^{IQ})$, with the rest of the maps zero
for $\alpha, \beta$ in this range.  
The corresponding RHS constants are $B_0 = E$ (where $E$ is
the all-ones matrix in $H(|S|)$), and $B_\alpha = 0$ 
($1 \le \alpha \le q$);  thus far we have imposed all the trace
constraints on query-updating (constraints $0,...,q-1$ give the 
effect of the pre-query unitary and query, while constraint
$q$ gives the effect of the unitary following the last query).
$\ca_{q+1, q}$ is 
minus the identity map, while $\ca_{q+1,q + z}$, for  
$z \in |T|$, is the identity map $\id: X \mapsto X$ (and the other maps $\ca_{q+1, x}$
are zero).  The corresponding RHS 
constants are zero:  this imposes the constraint that the $\Gamma_z$
are an additive decomposition of $\rho^{I}(q)$ into positive 
matrices.  
Finally, for $\alpha = q  + 1 + z, z \in |T|$, 
$\ca_{\alpha, q + z}: X \mapsto \Delta_z * X$, 
$\ca_{\alpha, q + 1 + z + |T|} = - {\rm id}$, and the
rest of them are zero.   And the corresponding RHS
constants, $B_\alpha: \alpha = q+1 +z, z \in T,$ are zero matrices.  These
just impose the output conditions, in the equality-constraint
form with slack variables given above.

To make this clearer, we display in Appendix B the constraints in the form 
$Ax = b$, where $A$ is the matrix of maps $A_{\alpha \beta}$, 
$x$ and $b$ are column vectors of  matrices $X_\alpha$, $B_\beta$;
we also display there the dual matrix-multiplication part of the dual 
constraints.  Appendix B serves as a useful aid to verifying that the
procedure about to be described for deriving the dual of
$P(g, \varepsilon, q)$ is  
carried out correctly, and that problem $D(g, \varepsilon, q)$ 
below is the result.

The dual feasible set is obtained, using 
Theorem 2, by transposing the 
matrix of maps, and replacing each map with its dual.  When 
$\ca : I \otimes Q \rightarrow I$ is the partial trace map, 
its dual $\ca^*: I \rightarrow I \otimes Q$ is given by 
$\ca^*: L \mapsto L \otimes I$ (where, to clear up ambiguous notation, 
$I$ in this
last specification refers to the identity matrix on the system
$Q$, not to the 
system $I$ itself as it does in the preceding two).
For $\ca: G  \mapsto \tr_Q (\Omega G \Omega^\dagger)$, we have 
$\ca^*: L \mapsto \Omega^\dagger (L \otimes I) \Omega$.  

\noindent
{\bf Remark:}
We can
give more explicit forms of these maps (and incorporate the special
form of $\Omega$, in the second case).  Viewing elements of
$I \otimes Q$ as block 
matrices blocked according to $X,Y \in S$, and elements of $I$
as matrices with elements indexed by pairs $X,Y \in S$, 
we have, when $\ca$ is the partial trace map, that $\ca^*$
takes $M$ to the matrix whose blocks are $M_{XY} I$.  For 
$\ca^{*}: G \mapsto \Omega^\dagger (M \otimes I) \Omega$, the
output matrix is the one whose blocks are $M_{XY} Y^\dagger X$.
Id is of course dual to itself, and so, as is easily verified, 
are the maps $M \mapsto \Delta_z * M$. \QED

We thus obtain a version of the dual program $D(g, \varepsilon, q)$.
The dual variables are 
$q + 1 + |T|$ $|S| \times |S|$ Hermitian matrices $Y_\beta$ whose
matrix elements are indexed by input-pairs $(X,Y) \in S \times S$.
The first $q$, corresponding to the primal query updating constraints,
we call $L_t,$ $(t \in \{0,...q-1\})$;  
the next, corresponding to the primal constraint that the
$\Gamma_z$ add up to $\rho^I(q)$, we call $L_q$; and the last $|T|$,
each corresponding to the output constraint on a primal 
variable $\Gamma_z$, we call
$\Lambda_z (z \in  T)$.   
We must find such matrices 
satisfying  the constraints:
\beqa
L_{(t-1)} \otimes I - \Omega^\dagger (L_{t} \otimes I) \Omega   \succeq 0
~~(1 \le t \le q) \\
L_q = L_{q+1} \\
L_{q+1}  \succeq - \Delta_z * \Lambda_{z+1}, ~~(1 \le z \le |T|) \\ 
-\Lambda_z \preceq 0 ~~(1 \le z \le |T|)  \\
\sum_{X,Y \in S} (L_0)_{X,Y} + (1 - \varepsilon)
\tr \sum_{z \in T} \Delta_z * \Lambda_z  < 0 \;.
\eeqa

Redefining the $\Lambda_z$ to be the negatives of the $\Lambda_z$ above, 
so as to 
have them be PSD, changing some signs, and dropping the redundant variable
$L_{q+1}$, we formally define the dual program:

\begin{definition} \label{def: dual SDP}
The semidefinite program (feasibility problem) $D(g, \varepsilon, q)$ is defined 
as the problem of finding $q+1  ~~|S| \times |S|$ Hermitian matrices $L_q, 
q \in \{0,...q\}$ and $|T|~~ |S| \times |S|$ Hermitian matrices $\Lambda_z$
for $z \in T$, with matrix elements indexed by $S \times S$, such that:
\beqa
L_{(t-1)} \otimes I \succeq  \Omega^\dagger (L_{t} \otimes I) \Omega^\dagger 
~~(1 \le t \le q) \label{first constraint in dual}\\
L_q  \succeq  \Delta_z * \Lambda_{z}, ~~(1 \le z \le |T|) \\ 
\Lambda_z \succeq 0 ~~(1 \le z \le |T|)  \\
\sum_{X,Y \in S} (L_0)_{X,Y} < (1 - \varepsilon) 
\sum_{z \in T} \sum_{X: g(X)=z} ~~(\Lambda_z)_{X,X} \;.
\eeqa
\end{definition}

Comparison to the program $\hat{P}(f,t,\varepsilon)$ of  Theorem 2 
in \cite{BSS2003} shows that they are identical except for the 
first constraint (the query-updating one), and that when $\Omega$ has
the special form corresponding to classical phase queries to 
input strings $x$ (when $x$ is in the input register), then
$D$ above specializes to $\hat{P}$ of \cite{BSS2003}.

Note that the constraint (\ref{first constraint in dual}) says that 
the block matrix whose $X,Y$ block is the $N \times N$ matrix 
$L_t[X,Y] X^\dagger Y  - L^{t-1}[X,Y] I$ is positive semidefinite.   

An immediate consequence of Theorem \ref{theorem: main characterization theorem}
and \ref{lemma: specialized duality lemma}, is the following Theorem.

\begin{theorem} \label{theorem: algorithms and duality}
With $S$, $T$ as above, a
$q$-query, $\varepsilon$-error quantum algorithm to compute
$g: S \rightarrow T$ exists if and only if a feasible solution 
to $D(g, q, \varepsilon)$ does not.
\end{theorem}

\section{Relaxation, duality, and a generalized spectral adversary method}

\subsection{Relaxation to the pairwise output condition: primal and dual programs}
We now consider relaxing the primal program by substituting the weaker output
condition of ``pairwise near-orthogonality,'' also known as  
the ``Ambainis condition'' \cite{Ambainis2000a}:
\beq
|\rho^I(q)[X,Y]| \le 2 \sqrt{\varepsilon(1 - \varepsilon)}
{\rm ~ when ~} g(X) \ne g(Y)\;.
\eeq 

We call it ``pairwise near-orthogonality'' because, 
by (\ref{eq: final-states gram matrix is input register density matrix}), 
when $\rho^{IQ}(q)$ is viewed as the unnormalized density matrix of the input
register in the explicit-inputs model, 
$|\rho^{IQ}(q)[X,Y]|$ is the modulus of the inner product of the $QW$ computer
states conditional on inputs $X$ and $Y$ in the ``black-box'' model, so it 
states that these conditional states are nearly (for small $\varepsilon$) orthogonal
if $X$ and $Y$ have different values of $g$; a necessary, but not sufficient, condition
for them to be the final states in a successful computation of $g$. 

In order to formulate this as a semidefinite constraint, we need constant
matrices $V^{XY} \in M(|S|)$, for all {\em unordered} pairs $(X,Y)$ of $X,Y \in S$ 
such that $g(X) \ne g(Y)$ (we call this set $\tilde{R}$ for future reference). 
For each such pair we define $V^{XY}$ to be 
the matrix whose $X,Y$ and $Y,X$ matrix elements are $1$, and whose
other matrix elements are all zero.  We also need the constant matrices
$W^{XY}$ for the same unordered input-pairs, but whose $X,X$ and $Y,Y$ matrix
elements are $1$ (and whose others are zero).   Then the 
Ambainis output condition
is equivalent to the conditions:
\beq
V^{XY}*\rho^I(q) + 2\sqrt{\varepsilon(1-\varepsilon)} W^{XY} \succeq 0\;
\eeq
where $(X,Y) \in R$.  We won't need the output variables $\Gamma_z$ 
in this case, but we will need a slack variable 
$\Pi^{XY} \succeq 0$ for each of the $|R|$ unordered pairs, to get equality constraints
\beq
V^{XY}*\rho^I(q) + 2\sqrt{\varepsilon(1-\varepsilon)} W^{XY} = \Pi^{XY}\;.
\eeq

Thus the dual program is to find $|S| \times |S|$ Hermitian matrices 
$L^t$, $0 \le t, \le q$ and $\Upsilon_{XY}$, $X, Y \in S$, such that:
\beqa
L_{t-1} \otimes I \succeq  \Omega( L_{t} \otimes I) \Omega^\dagger
~(1 \le t \le q) \\
 L_{q} 
\succeq \sum_{(X,Y) \in R} (V^{XY} * \Upsilon_{XY}) 
\\
\Upsilon_{X,Y} \succeq {\bf 0}~ (X,Y) \in R)
\\
 \sum_{X,Y \in S} L_0[{X,Y}] < - 2 \sqrt{\varepsilon(1 - \varepsilon)}
\sum_{(X,Y) \in R} \tr(\Upsilon^{XY} W^{XY})\;,
\eeqa
and $(\Upsilon_{XY})_{MN} = 0$ unless $MN \in \{XX, XY, YX, YY\}$.

Rewriting this in terms of the variables $K_t := - L_{q-t}$ we
formally define the dual program $D_A$.

\begin{definition}\label{def: dual of relaxed SDP}
\beqa
K_0 \preceq 
- \sum_{(X,Y) \in R} (V^{XY} * \Upsilon_{XY})  
\label{special constraint on K0}
\\ 
K_{t-1} \otimes I \preceq 
\Omega (K_{t} \otimes I) \Omega^\dagger 
~(1 \le t \le q) \label{relaxed dual query constraint} \\ 
\label{constraint: upsilons psd}
\Upsilon_{X,Y} \succeq {\bf 0}~ ((X,Y) \in R)
\\ \label{africabrass}
 \sum_{X,Y \in S} K_q[X,Y] > 2 \sqrt{\varepsilon(1 - \varepsilon)}
\sum_{(X,Y) \in R} \tr \Upsilon^{XY} \;,
\eeqa
and $(\Upsilon_{XY})_{MN} = 0$ unless $MN \in \{XX, XY, YX, YY\}$.
\end{definition}

\subsection{A generalized spectral adversary method}

We next obtain, from this dual program, a generalization
of Theorem 4 of \cite{BSS2003}, giving a lower bound directly on the
number of queries in an algorithm $\varepsilon$-computing a function, in
terms of relatively easily computed properties of the function and a 
``weight matrix'' $\Gamma$ that we are free to choose.  This
gives a generalization of the so-called ``spectral adversary
method'' for quantum query complexity lower bounds.  We use the
notation $\lambda(M)$ for the largest eigenvalue of a matrix $M$.

\begin{theorem} Let $S$ be a finite set of unitary $|S| \times |S|$ matrices, 
and let $g: S \mapsto T$, $T$ a finite set.
Let $\Gamma$ be a nonnegative real symmetric $|S| \times |S|$ matrix indexed
by $S$, such that $\Gamma_{X,Y} = 0$ whenever $g(X) = g(Y)$.
Then
\beq
QQC_{\varepsilon}(g) \ge 
\frac{
(1 - 2 \sqrt{\varepsilon(1 - \varepsilon)})
\lambda(\Gamma)
}{
2 \lambda( \Gamma \otimes I - \Omega ( \Gamma \otimes I) \Omega^\dagger)
}
\;.
\eeq
\end{theorem}
\begin{proof}
To prove this, we construct, for any $\Gamma$ as above and $q$ below
the bound given in the theorem, a sequence $K_t : 0 \le t \le q, 
\Upsilon_{XY}, \{X,Y\} \in R$
that is a feasible solution to $D_A(g, q, \varepsilon)$.

Note that by the standard Perron-Frobenius theory of nonnegative
matrices \cite{Horn85a}, $\Gamma$ has a normalized
eigenvector $v$  with nonnegative entries, whose eigenvalue is 
$\Gamma$'s largest, i.e. $\lambda(\Gamma)$.  
We define $K_t := ( \Gamma - t \alpha I)*vv^t$, 
where $\alpha := 
2 \lambda( \Gamma \otimes I - \Omega ( \Gamma \otimes I) \Omega^\dagger)$.
We also define $\Upsilon_{XY}$ via 
\beqa
\Upsilon_{XY}[X,X] = \Upsilon_{XY}[Y, Y] 
= -\Upsilon_{XY}[X,Y] = -\Upsilon_{XY}[Y,X] := \nonumber \\
K_0[X,Y] \equiv \Gamma[X,Y]v[X]v[Y]\;,
\eeqa 
with its other matrix elements zero.  These are manifestly positive semidefinite,
satsfying (\ref{constraint: upsilons psd}). 
That (\ref{special constraint on K0}) is satisfied with equality is also immediate
from the definitions. 

To verify that (\ref{relaxed dual query constraint}) is satisfied, we 
have a look at
\beqa
\Omega(K_t \otimes I)\Omega^\dagger - K_{t-1} \otimes I = 
\Omega((\Gamma - t\alpha I)*vv^t)\otimes I) \Omega^\dagger
-((\Gamma - (t-1) \alpha I)*vv^t) \otimes I \\
= (vv^t\otimes E) * \Omega(\Gamma - t \alpha I) \otimes I) \Omega^\dagger
- (vv^t \otimes E)*((\Gamma - (t-1) \alpha I) \otimes I)  \\ 
= (vv^t \otimes E)* \left[\Omega(\Gamma \otimes I) \Omega^\dagger
- t \alpha (I \otimes I) - \Gamma \otimes I + (t-1) \alpha(I \otimes I) \right] \\
= (vv^t\otimes E)* \left[ \Omega(\Gamma \otimes I) \Omega^\dagger - \Gamma \otimes I
- \alpha(I \otimes I) \right]
 \label{last line}\;.
\eeqa
Note that in the second equality we used the identity
\beq
Z^\dagger( X * M \otimes I) Z 
\equiv (M \otimes E)*Z^\dagger(X \otimes I)Z\;,
\eeq
which does not hold for general $Z$, but does hold when
(as in the cases $Z = \Omega, Z = I$ that we use)
$Z$ is block-diagonal when the blocks are indexed by a 
basis for the input register (the register that we write on the left 
in tensor products).  
The matrix in (\ref{last line}) is positive 
semidefinite by the definition of $\alpha$, so the constraint
(\ref{relaxed dual query constraint}) is indeed satisfied.
Finally, the constraint (\ref{africabrass}) is 
satisfied because 
\beq
\sum_{(X,Y) \in R} \tr \Upsilon_{XY} 
=  \sum_{(X,Y) \in R} 2 \Gamma[X,Y]v[X]v[Y]  
= v^t \Gamma v = \lambda(\Gamma)\;,
\eeq 
while $\sum_{XY} K_q[X,Y] = \lambda(\Gamma) - q \alpha$, which by our assumption on
$q$ is greater than or equal to $2 \sqrt{\varepsilon(1 - \varepsilon)}
\lambda(\Gamma)$.  
\QED
\end{proof} 

It is easily seen that this Theorem specializes to Theorem 4 of 
\cite{BSS2003}.

\acknowledgments
We thank the DOE and NSF for support.
\begin{appendix}
\section{The matrix multiplications appearing 
in the primal and dual constraints}

In this section we use the notation $\trq$ to denote the partial trace
map from $I \otimes Q$ to $I$, $\Omega$ to denote the map $G \mapsto
\Omega G \Omega^\dagger$, $\Delta_k*$ to denote the map $G \mapsto
\Delta_k * G$ (where $*$ is the elementwise matrix product);
juxtaposition of maps to indicate composition (thus $\trq \Omega: G
\mapsto \tr_Q(\Omega G \Omega^\dagger)$, and the superscript $^*$ to
indicate the dual map.  We also use the facts that the maps
$\Delta_k*$ are self-dual and that the dual $\Omega^*$ of the map
$\Omega$ is the map $\Omega^\dagger: M \mapsto \Omega^\dagger M
\Omega$.

\subsection{Unrelaxed constraints}
With this notation, 
the matrix multiplication portion of the primal constraints is:

\beq
\left(
\begin{array}{cccc|cccc|ccccc}
\trq & & & &   & &  & & & \\
-\trq \Omega  & \trq  & & 
& &  & & & &  \\
& \ddots & \ddots & 
& &  & & & &  \\
&   & -\trq \Omega & \id 
& &  & & & & \\
\hline 
&   &   & -\id
& \id & \id & \cdots & \id  & & & \\

\hline
&   & &  &  
\Delta_1* &  &    & & & -\id &  & \\
&   &   &  
 & & \Delta_2*  &    & &   & & -\id & \\
&   &     
&  &   &  & \ddots & &  & &  & \ddots &  \\
&   &   &  
 & &   &   &  \Delta_{|T|}*  & & & & & -\id 
 \end{array}
\right)
\left[
\begin{array}{c}
\rho^{IQ}(0) \\ \rho^{IQ}(1) \\ \vdots \\ 
\rho^{IQ}(q-1) \\ \hline \rho^{I}(q) \\ \hline 
\Gamma_1 \\
\Gamma_2 \\
\vdots \\
\Gamma_{|T|} \\ \hline  \Pi_1 \\ \vdots \\ \Pi_{|T|} 
\end{array}
\right]
=
\left[
\begin{array}{c}
E \\ {\bf 0} \\ \vdots \\ {\bf 0} \\  \hline 
 {\bf 0}   
\\  \hline  (1 - \varepsilon) \Delta_1
\\ (1 - \varepsilon) \Delta_2 
\\ \vdots \\
(1 - \varepsilon) \Delta_{|T|}  
\end{array}
\right]
\eeq

The matrix multiplication part of the dual constraints is:

\beq
\left(
\begin{array}{ccccc|c|ccc}
\trq^* & - \Omega^\dagger \trq^*  & & & &  & & & \\
  & \trq * & -\Omega^\dagger \trq^* 
& &  & &  & & \\
& & \ddots & \ddots & 
& & & & \\
&  & & \trq^* & - \Omega^\dagger \trq^* &
& & &  \\
& & & &  \id & -\id & & & \\
\hline
& & & & & \id & \Delta_1* & & \\
& & & & & \vdots & & \ddots   & \\  
& & & & & \id & & &  \Delta_{|T|}*  \\
\hline 
& & & & &  & -\id & & \\
& & & & &  & & \ddots   & \\  
& & & & &  & & &  -\id  
\end{array}
\right)
\left[
\begin{array}{c}
L_0 \\ L(1) \\ \vdots \\ L_q \\
\hline L(q+1) \\ \hline 
\Lambda_1 \\
\Lambda_2 \\
\vdots \\
\Lambda_{|T|} 
\end{array}
\right]
\succeq 
\left[ 
\begin{array}{c}
{\bf 0} \\  \vdots \\ \0 \\ \hline  
 {\bf 0}  \\ \vdots \\ \0  \\ \hline \0 \\ \vdots \\ \0 
\end{array}
\right] 
\eeq

\subsection{Relaxed constraints (pairwise output condition)}

Primal matrix multiplication constraints:

\beq
\left(
\begin{array}{cccc|c|ccccc}
\trq & & & &   & & & &  \\
-\trq \Omega  & \trq  & & 
& &  & & &  & \\
& \ddots & \ddots & 
& &  & & &   & \\
&   & -\trq \Omega & \trq  
& &  & & & &   \\
&   &  & -\trq \Omega 
&\id &  & & &  & \\
\hline &  & & &   -V_{X_1,Y_1}*  &  
 \id &  &  \\
&   & &   & 
  -V_{X_2,Y_2}*  & &  \id & \\
&   &    & 
&  \vdots &   &  &  \ddots &  \\
&   &   &  &
 & &   &   & \id 
 \end{array}
\right)
\left[
\begin{array}{c}
\rho^{IQ}(0) \\ \rho^{IQ}(1) \\ \vdots \\ 
\rho^{IQ}(q) \\ \hline \rho^{I}(q) 
\\ \hline 
\Pi_{X_1,Y_1} \\ \Pi_{X_2,Y_2}\\ \vdots \\ \Pi_{X_{|R|},Y_{|R|}} 
\end{array}
\right]
=
\left[
\begin{array}{c}
E \\ {\bf 0} \\ \vdots \\ {\bf 0} \\  {\bf 0} \\   
\\  \hline  2 \sqrt{\varepsilon(1 - \varepsilon)} W_{X_1,Y_1}
\\ 2 \sqrt{\varepsilon(1 - \varepsilon)} W_{X_2,Y_2} 
\\ \vdots \\
2 \sqrt{\varepsilon(1 - \varepsilon)} W_{X_{|R|},Y_{|R|}}  
\end{array}
\right]
\eeq

From the above we get the dual matrix multiplication constraints:

\beq
\left(
\begin{array}{ccccc|ccc}
\trq^* & -\Omega^\dagger  \trq^*  & & &   & & & \\
  & \trq^* & -\Omega^\dagger \trq^* 
& &  & &  &  \\
& & \ddots & \ddots  
& & & & \\
&  & & \trq^* & - \Omega^\dagger \trq^* &
 & &  \\ \hline
&  & & &  \id & - V_{X_1,Y_1}*  & \cdots  & 
- V_{X_{|R|},Y_{|R|}}*\\
\hline
& & & & &  \id & & \\
& & & & &   & \ddots   & \\  
& & & & &  & & \id  
\end{array}
\right)
\left[
\begin{array}{c}
L_0 \\ L(1) \\ \vdots \\ L_q \\
\hline 
\Upsilon_1 \\
\Upsilon_2 \\
\vdots \\
\Upsilon_{X_{|R|},Y_{|R|}} 
\end{array}
\right]
\succeq 
\left[
\begin{array}{c}
{\bf 0} \\  \vdots \\ \0 \\ \hline \0 \\ \hline \0 \\ \vdots \\ \0 
\end{array}
\right]
\eeq

\end{appendix}

\vspace*{-5mm}


\end{document}